\documentclass[fleqn]{article}
\usepackage{amssymb}
\author{\normalsize Sergey S. Kokarev\thanks{email: sergey@yspu.yar.ru}\\
\small\it Department of theoretical physics, r.409, Yaroslavl State Pedagogical
University,\\ \small\it Respublikanskaya 108, Yaroslavl, 150000, Russia}
\title{\large\bf Gravity as a bend of 4D elastic plate}
\date{}
 \DeclareSymbolFont{AMSa}{U}{msa}{m}{n}
\DeclareMathSymbol\square{\mathord}{AMSa}{"03} \global\let\Box\square

\begin{document}
\maketitle

\begin{abstract}
Gravity is represented as a linear theory of  strong plate bend at the
variational functionals level. Some estimates of elastic constant of space-time
are made. Physical (elastic) sense of field lagrangians and Einstein Equations
is discussed.
\end{abstract}

The question about physical nature of space, time, gravity and Einstein
Equations (EE) has been discussing   from the earliest era of special (SR) and
general (GR) relativity up to a present time. Today we have powerful
mathematical means of formulation and investigation of space-time physics,
while physical foundations of the theories and their relations both to
observable world and to other physical topics often remain beyond the scope of
attention.

In my report I am going to demonstrate physical relevance of linear elasticity
theory (Hooks law) language for formulation and clarifying standard GR. Note,
that difficulties of most of the  attempts, that  have been made in this
direction, were caused by restriction to (at best) 4D
elasticity(\cite{born}-\cite{tart})
--- we'll see that it totally excludes the possibility of "embedding" gravity
into elasticity.

The starting points of our consideration are two quite  obvious observations:

1) We locally live in  {\it 4D Minkowski world} $\mathbb{M}_4$;

2) In every 3D simultaneity section, which is locally isometric to 3D euclidian
space $\mathbb{E}_3,$ we  deal with (approximate) {\it Hooks law,} when
investigate contact interactions properties of real mechanical bodies:
\begin{equation}\label{hook3}
\sigma=2\mu\mathcal{D}+\lambda{\rm Tr}[\mathcal{D}]\eta,
%=
%\frac{E}{1+\sigma}\left(\mathcal{D}+\frac{\sigma}{1-2\sigma}{\rm Tr
%}[\mathcal{D}]{\eta}\right),
\end{equation}
where $\sigma$ and $\mathcal{D}$ --- 3D {\it stress} and {\it strain tensors,}
defined by expressions\footnote{I'll  use indexless representation.
Particularly, $\partial$ and $\partial^2$ will denote partial derivative
operators of
 first and second orders respectively.}:
\begin{equation}\label{strstr}\mathcal{D}\equiv\frac{1}{2}\left[(\overrightarrow\partial\otimes\xi+\xi\otimes\overleftarrow\partial)
+(\partial\otimes\partial)\eta(\xi,\xi)\right];\ \ \
\sigma=\frac{\partial\mathfrak{F}}{\partial\mathcal{D}},
\end{equation}
 $\xi=\xi(x)$ --- {\it displacements vector field}, $\mathfrak{F}$
--- {\it elastic energy density}:
\begin{equation}\label{fren}
\mathfrak{F}=\mu\mathcal{D}^2+\frac{\lambda}{2}{\rm Tr}^2[{\mathcal{D}}],
\end{equation}
 $\mu,\lambda$ ---
pair of independent elastic constants
--- {\it Lame coefficients}, $\eta$
--- euclidian metric in $\mathbb{E}_3$ \cite[\S4]{land1}.
There are many different generalizations of these two points that can be
assumed to reveal physical properties of space, time and gravity. Now we accept
the most simple ones (\cite{kok1}-\cite{kok2}):

1) We live on {\it pseudoeuclidian 4D plate,} i.e. such multidimensional body
$\mathcal{P}$, whose sizes along some $\mathbb{M}_4$ are much larger, than
along orthogonal to $\mathbb{M}_4$ dimensions: \[\mathcal{P}\simeq
\mathbb{M}_4\times\overline{\mathbb{M}}_{N}\simeq\mathbb{M}_4
\mathop{\times}\limits_{i=1}^N\left[-\frac{h_i}{2};\frac{h_i}{2}\right]\subset\mathbb{M}_{N+4},\]
where "$\simeq$" denotes homeomorphism relation, $\{h_i\}$ $(i=\overline{1,N})$
--- the set of small thicknesses of the plate in extradimensions,
$\mathbb{M}_{N+4}$ --- embedding space-time.

2) This 4D plate possesses isotropic elastic properties, which we'll describe
by the same denoted pair of Lame coefficients $\lambda$ and $\mu$ in
$N+4$-dimensional Hooks law:
\begin{equation}\label{hookmtd}
\sigma=2\mu\mathcal{D}+\lambda{\rm Tr}[\mathcal{D}]\Theta,
%\frac{E}{1+\sigma}\left(\mathcal{D} +\frac{\sigma}{1-(N+3)
%\sigma}{\rm Tr
%}[\mathcal{D}]\Theta\right),
\end{equation}
that generalizes (\ref{hook3}) ($\Theta=\theta+\bar{\theta}$ in (\ref{hookmtd})
--- metrics on  $\mathbb{M}_{N+4}$, $\mathbb{M}_4$ and $\overline{\mathbb{M}}_N$
correspondingly and all other notations are related to the multidimensional
plate). The (\ref{hookmtd}) determines multidimensional elastic energy density,
having the same form as in (\ref{fren}), which we intend to calculate.

Imagine that the plate in unstrained state is placed in $\mathbb{M}_{N+4}$,
occupying a subspace defined by the equation: $\bar x=0$, where $\{\bar x\}$
together with $\{x\}$ form $(N+4)$-dimensional Cartesian coordinate system:
$\{X\}=\{x\}\times\{\bar x\},$ adjusted to the plate surface $\mathbb{M}_4$ (so
$\partial_{x}\in T\mathbb{M}_4,\
\partial_{\bar x}\in T\mathbb{M}_{N}$). Then its mechanical straining
can be described by a smooth displacement vector field  $\Xi$:
$\mathcal{P}\stackrel{\scriptstyle\Xi}{\to}\mathcal{P}',$ that takes in our
coordinate frame the form
%\begin{equation}\label{def}
$X\to X'=X+\Xi$ or in projections $x'=x+\xi;\ \ \bar{x}'=\bar{x}+\bar\xi,$
%\end{equation}
where $P(X)\in\mathcal{P}\to P(X')\in\mathcal{P}'$ --- some arbitrary matter
point inside the body, $\Xi=\xi+\bar{\xi}$, $\xi\in T\mathbb{M}_4$, $\bar\xi\in
T\mathbb{M}_N.$

It is well known, that within linear elasticity theory deformations normal and
tangent to the plate are factorizable: their contributions into elastic energy
density can be calculated separately and independently \cite[\S11,\S13]{land1}.
Let us start from the

{\bf 1. Bend deformation.} We have pure bend without stretching with  $\xi=0,\
\partial^2\bar\xi\neq0.$ Suppose, that middle (in thicknesses) plane
$\mathbb{M}_4$ of the unstrained plate is transformed by the bending $\bar\xi$
into some Riemannian manifold $\mathbb{V}_4.$
% (see fig.\ref{defor}).
This new middle surface in case of a weak bend is usually called {\it neutral
surface} because tangent stresses $\sigma|_{\mathbb{V}_4}=0$.
%\begin{figure}[htb]
%\centering \unitlength=0.5mm \special{em:linewidth 0.4pt} \linethickness{0.4pt}
%\begin{picture}(139.00,60.00)(0.00,50.00)
%\bezier{476}(23.00,92.00)(71.00,61.00)(125.00,92.00)
%\bezier{452}(21.00,73.00)(71.00,51.00)(125.00,73.00)
%\put(23.00,92.00){\line(-1,-6){3.00}} \put(125.00,92.00){\line(0,-1){19.00}}
%\put(9.00,70.00){\vector(1,0){130.00}} \put(72.00,70.00){\vector(0,1){36.00}}
%\put(23.00,81.00){\line(5,-2){5.00}} \put(32.00,77.00){\line(3,-1){6.00}}
%\put(43.00,74.00){\line(4,-1){7.00}} \put(54.00,71.00){\line(6,-1){6.00}}
%\put(81.00,70.00){\line(6,1){9.00}} \put(95.00,72.00){\line(6,1){8.00}}
%\put(108.00,74.00){\line(5,2){7.00}} \put(119.00,78.00){\line(2,1){6.00}}
%\put(132.00,63.00){\makebox(0,0)[cc]{$x$}}
%\put(65.00,100.00){\makebox(0,0)[cc]{$\bar x$}}
%\put(72.00,66.00){\makebox(0,0)[cc]{O}} \put(115.00,70.00){\vector(0,1){7.00}}
%\put(115.00,64.00){\makebox(0,0)[cc]{$\bar\xi$}}
%\put(30.00,100.00){\makebox(0,0)[cc]{$\mathbb{M}_{N+4}$}}
%\end{picture}
%\caption{Straining of a thin plate}\label{defor}
%\end{figure}
Since  both  thicknesses and bend are small we can omit  any bending external
force densities  in comparison with internal stresses (similarly to ordinary
2-D plates in 3D euclidian space  \cite[\S11]{land1}) and also omit variations
of $\bar{\partial}$: $\bar\partial'\approx\bar\partial$, where
$\bar\partial\equiv\partial_{\bar x}.$ It implies $\sigma(\bar\partial,\
)\approx0 $ at every point of $\mathcal{P}$. Then from (\ref{hookmtd}) we get
the following algebraic equations for strain tensor (and differential equations
for displacements field):
\[\mathcal{D}(\bar{\partial}_i,\bar{\partial}_j)=\mathcal{D}(\partial,\bar{\partial})=0;\
\ \ 2\mu\mathcal{D}(\bar{\partial}_i,\bar{\partial}_i)+\epsilon_i\lambda{\rm Tr
}[\mathcal{D}]=0,
\]
where $\epsilon_{i}=\pm1$ if $\bar{\partial}_i$ is time-like or space-like
respectively. Its solution is $\xi=-\bar{\theta}(\bar{x},\partial\bar\xi).$
Then for $\mathcal{D}$ we obtain $(N\lambda+2\mu\neq 0)$:
\begin{equation}\label{sol}
\mathcal{D}(\partial,\partial)=-\bar{\theta}(\bar x,\partial^2\bar\xi);\ \
\mathcal{D}(\bar{\partial}_i,\bar{\partial}_i)=-\epsilon_{i}\frac{\lambda
S}{N\lambda+2\mu};\ \  S\equiv{\rm div}\, \xi\equiv\theta(\partial, \xi)=
-\bar{\theta}(\bar{x},\Box\bar{\xi}).
\end{equation}
Substituting (\ref{sol}) into (\ref{fren}) we get
\begin{equation}\label{freepl}
\mathfrak{F}_{\rm b}=\mu \left\{\bar{\theta}^2(\bar{x},\partial^2\bar\xi)+
f\,\bar{\theta}^2(\bar{x},\Box\bar\xi)\right\},
\end{equation}
where factor $f=\lambda/(N\lambda+2\mu)$.

To obtain expression for total bend energy of the plate one should integrate
(\ref{freepl}) over its $N+4$-dimensional volume:
\begin{equation}\label{freegen}
 F_{\rm b}=\frac{\mu
H_{N}}{12}\int\limits_{\mathbb{M}_4}\left\{\bar{\theta}_{\rm
h}(\partial^2\bar{\xi} ,\partial^2\bar\xi)+ f\,\bar{\theta}_{\rm
h}(\Box\bar\xi,\Box\bar\xi)\right\}d{\rm m}[\mathbb{M}_4],
\end{equation}
where measure $d{\rm m}[\mathbb{M}_4]$ on $\mathbb{M}_4$ is defined from the
decomposition  $d{\rm  m}[\mathbb{M}_{N+4}]=d{\rm  m}[\mathbb{M}_{4}]\cdot
d{\rm m}[\mathbb{M}_{N}],$
\[
\frac{\bar{{\delta}}_{\rm h} H_{N}}{12}=
\int\limits_{\mathbb{M}_N}(\bar{x}\otimes\bar{x})\, d{\rm m}[\mathbb{M}_N].
\]
 The above equation has been used with $H_{N}\equiv\prod\limits_{i=1}^Nh_i$, $\bar{{\delta}}_h=
{\rm diag}(h_1,$ $\dots,h_N),$ $\bar{\theta}_{\rm h}\equiv\bar{{\delta}}_{\rm
h}\cdot\bar{\theta}={\rm diag}(\epsilon_1h_1,\dots,\epsilon_Nh_N).$  From the
view point of GR, expression (\ref{freegen}) should be relevant to gravity far
from its source, where bending is weak. Near the sources we need to take into
account tangent stretches and shears of the plate (i.e. energy-momentum tensors
of the sources) even within the linear elasticity.

{\bf 2. 4D stretches  and shears.} 4D elastic shears and stretches energy, that
is generated by a strong bend, has the form: \begin{equation}\label{str}F_{\rm
s}=H_N\int\limits_{\mathbb{M}_4}\mathfrak{F}_{\rm s}\,d{\rm m}[\mathbb{M}_4]
=\frac{H_N}{2}\int\limits_{\mathbb{M}_4}(\sigma\cdot\mathcal{D})\,d{\rm
m}[\mathbb{M}_4],
\end{equation}
where the standard representation of elastic energy density of tangent
deformation in linear elasticity theory \cite{land1}: $\mathfrak{F}_{\rm
s}=(\sigma\cdot\mathcal{D})/2=\sigma_{\alpha\beta}\mathcal{D}^{\alpha\beta}/2$
$(\alpha,\beta=\overline{0,3})$ has been used. So,  full variational functional
$F$, including potential energy of bending and stretching multidimensional
forces external to the plate, takes the form
\begin{equation}\label{var}
F=F_{\rm b}+F_{\rm s}+U.
\end{equation}

{\bf 3. Comparing with Einstein-Gilbert action of GR.} Now we compare action
(\ref{var}) with full action $S=S_{\rm g}+S_{\rm m}$ of a system "gravitational
field + matter source" in GR. To reveal the similarity of structures of $F_{\rm
b}$ and $S_{\rm g}$ we need to reformulate the latter in terms of embedding
theory in deformation representation, where  any Riemannian manifolds
$\mathbb{V}_4$ is interpreted as Minkowski plane $\mathbb{M}_4$, deformed by
some normal to the plate vector field $\bar\xi(x).$ Induced Riemannian metric
$g$  will then be
\begin{equation}\label{metr}
g=\theta+2\mathcal{D},
\end{equation}
%where $\theta$ --- flat Minkowski metric on undeformed $\mathbb{M}_4,$
%$\mathcal{D}=(\partial'\otimes\partial'')\bar\theta(\bar\xi',\bar\xi'')$ --- "strain
%tensor", $\bar{\theta}$ --- flat metric  of orthogonal to $\mathbb{M}_4$
%addition, accents denote arguments of partial derivatives.
where $\mathcal{D}$ is the second (nonlinear) term in (\ref{strstr}) and GR
action in this representation is: \begin{equation}\label{act} S_{\rm
g}=-\frac{c^{3}}{16\pi G}\int\limits_{\mathbb{V}_4} {}^{(4)}\mathcal{R}\, d{\rm
m}[\mathbb{V}_4]=-\frac{c^{3}}{16\pi G}\int\limits_{\mathbb{V}_4}\left\{
\mathcal{H}(\Box\bar\xi,\Box\bar\xi)-
\mathcal{H}(\partial^2\bar\xi,\partial^2\bar\xi)\right\}\, d{\rm
m}[\mathbb{V}_4],
\end{equation}
where ${}^{(4)}\mathcal{R}$ --- scalar curvature of $\mathbb{V}_4$,
$\mathcal{H}\stackrel{\rm def}{\equiv}\sum_{m=1}^{N}\epsilon_{m}\cdot$
$n_{(m)}\otimes n_{(m)}$ --- projector on subspace of $T\mathbb{M}_{N+4}$
orthogonal to $\mathbb{V}_4$, $\{n_{(m)}\}$
--- basis vector fields of the subspace, normalized by conditions:
$\Theta(n_{(m)},n_{(l)})=(\bar{\theta}_{\rm h})_{ml}$. Note, that under the
bend in linear approximation (when Hooks law is valid), we have
$\mathcal{H}\approx\bar{\theta}_h$, and $d{\rm m}[\mathbb{V}_4]\approx d{\rm
m}[\mathbb{M}_4]$ Then under $f=-1$ expressions (\ref{freegen}) for $F_{\rm b}$
and (\ref{act}) for $S_{\rm g}$ become identical up to a dimensional constant.
The remaining parts
--- $F_{\rm s }$ and $S_{\rm m}$ should be also identified, as those involving
tangent to $\mathbb{V}_4$ stresses (energy-momentum tensors). There are  no
analogies of external energy $U$ in standard physics, since it concerns
noncausal (from the viewpoint of $\mathbb{V}_4$) interaction of the plate with
its multidimensional environment.

{\bf 4. Discussion.}

Let us briefly discuss some general consequences of the approach.

{\bf Multidimensional elastic constants.}
%The similar structure of variational
%functionals (\ref{freegen}) and (\ref{act}) suggests that space-time in GR can
%be really viewed as 4D elastic plate with some specific properties.
It is easily to show that value of elastic parameter $f=-1$, reproducing
integrand of (\ref{act}) in (\ref{freegen}),  corresponds to the Poisson
coefficient of the plate medium $\sigma=1/2$ (\cite{kok1}). Second independent
elastic constant
--- Young modulus $E$ --- can be evaluated by  some dimensional
manipulations. The result is
\[
Eh^{N+3}\sim\frac{c^4}{G}\sim\frac{1}{\ae},
\]
that supports old Sacharov's considerations on possible relation between
Einstein constant and elastic properties of space-time \cite{sah}.
%  Assuming that extradimensions have an order of the Planck length
%$h_{\rm Pl}=\sqrt{G\hbar/c^{3}}\approx10^{-33}$cm, we obtain  $E\sim
%10^{144+35N}\mbox{Pa}$ or effective 4D Young modulus $E_{\rm eff
%}=Eh^N\sim10^{144}$Pa. We see that space-time plate in this case possesses huge
%stiffness, which can be responsible for observable "flatness" of space-time in
%any local region.

{\bf Lagrange formalism as 4-D elasticity theory.} Let us draw attention to
analogy between these two expressions: \[ \delta F_{\rm
s}=\int\sigma\cdot\delta{\cal D}\, d{\rm m};\ \ \delta S_{\rm m
}=\frac{1}{2c}\int \mathcal{T}\cdot\delta g\, d{\rm m},\] where $d{\rm m}$
denotes suitable form of volume of the base manifold. The first is general
thermodynamic relation, connecting stress tensor with infinitisemal variation
of an elastic free energy \cite[\S3]{land1}. The second
--- is well known rule for calculation of symmetric energy-momentum tensor $\mathcal{T}$ of
fields or matter from its lagrangian density $\mathcal{L}$, $S_{\rm m}[q]=
\int\mathcal{L}(q,\partial q )\, d{\rm m}$, where $\{q\}$ and $\{\partial q\}$
are sets of field variables and their derivatives respectively \cite{land2}.
Our approach shows, that these expressions have deep interrelations, under
assumption of the relation $F_{\rm s}=cS_{\rm m}.$ Really, curvelinear metric
$g$ can be understood as result of tangent to $\mathbb{M}_4$ diffeomorphism
 $x\stackrel{\scriptstyle \xi}{\to} x'=x'(x)$,
which has not passive (as in GR), but active sense. Then  in accordance with
(\ref{metr}) metric on $\mathbb{M}_4'$ will be $ g=\theta+2{\cal D},$ where
$\theta$ --- metric on original $\mathbb{M}_4.$ Both $\mathfrak{F}_{\rm s}$ and
$\mathcal{L}$ implicitly contain metric $g$ (to get scalar expressions from
$\cal D$ or $q,$ and $\partial q$). From the kind of $g$ we get $\delta
g=2\delta{\cal D},$
and so
\begin{equation}\label{iden1}
\sigma=\frac{\delta F_{\rm s}}{\delta{\cal D}}=2\frac{\delta F_{\rm s}}{\delta
g}=\frac{\delta S_{\rm m}}{\delta g}=\mathcal{T}.
\end{equation}
If this analogy is not occasional, then {\it any classical field lagrangian can
be interpreted as elastic energy density of some strain of $\mathbb{M}_4$ and
specific choice of field variables is determined by kind of the straining:
 $\mathcal{D}=\mathcal{D}(q)$} \cite{kok2}.

{\bf Physical essense of Einstein Equations.} Extremality of $F[\Xi]$ and
$S[q]$ leads to Euler-Lagrange equations, which in turn, provide validity of
equilibrium equations in the first case and conservation laws in the second:
\begin{equation}\label{conserv} \delta F=0\ \longrightarrow \ {\rm div}\, \sigma=0;\ \ \ \
\delta S=0\ \longrightarrow \ {\rm div}\, T=0.\
\end{equation}
Within the present approach it would be natural to use unified language and
regard conservation laws  as equilibrium equation of some elastic body. In view
of presented results, we conclude, that  this  body is {\it space-time itself.}

Assume, that space-time plate is characterized by "phenomenological"
multidimensional elastic constant $E$ and $\sigma$:  $S_{\rm g}=S_{\rm
g}(E,\sigma)$, where $S_{\rm g}$ --- is generalized action for gravitational
field, or free elastic energy of bending (\ref{freegen}). Then
\begin{equation}\label{varS}
\delta_g S=\int(\mathcal{T}^{(n)}(E,\sigma)+\mathcal{T}^{(t)}) \cdot\delta g\,
d\,{\rm m}.
\end{equation}
where $\mathcal{T}^{(n)}$ is generated by  normal straining of the plate, the
$\mathcal{T}^{(t)}$ --- by  tangent ones. Vanishing of (\ref{varS}) gives
\begin{equation}\label{non}
\mathcal{T}^{(n)}(E,\sigma)+\mathcal{T}^{(t)}=0
\end{equation}
--- "generalized" Einstein equations.
One can conclude, that Einstein theory, even in this "generalized" variant,
operates with {\it nonstressed} state of space-time. In other words, {\it
physical meaning of standard Einstein equations is expressed in
intercompensation of tangent stres\-ses, generated by normal and tangent plate
strains.} From the view point of elasticity true dynamical variables are
components of strain vector $\Xi$ and true equilibrium equations of the plate
are:
\begin{equation}\label{eqein}
{\rm div}(\mathcal{T}^{(n)}(E,\sigma)+\mathcal{T}^{(t)})=0.
\end{equation}
The (\ref{eqein}) are consequence of {\it the strong bend} equations:
\[\Box^2\bar\xi_{D}-H_N\theta(\partial,\sigma(\ ,\partial\bar\xi))=\bar P;\ \
{\rm and}\ \  H_N\,{\rm div}\sigma=-P,\] which are deduced by variational
procudure over $\Xi$ and so are true (necessary) equilibrium equations. Here
$\{D\}$
--- the set of {\it cylindrical stiffness} factors of the plate  in
extradimensions, $\Pi=P+\bar P$
--- total, tangent, and normal to the plate surface multidimensional force
densities respectively \cite{kok2}. As it has been mentioned earlier, one can
go to Einstein GR by setting $\sigma=1/2$. Under this condition
$\mathcal{T}^{(n)}(E,1/2)$ transforms to
\[-G/\ae=-({}^{(4)}{\cal R}{\rm ic}-(1/2)g\,{}^{(4)}{\cal R})/\ae\]
--- purely geometrical Einstein tensor, which satisfies ${\rm div}\, G\equiv0$ because of Bianchi
identities! Then from (\ref{eqein}) automatically follows ${\rm div}\,
\mathcal{T}^{(t)}\equiv0$ and we obtain well known statement: "motion equations
are contained in the field equations of GR".
%This  fundamental property of GR
%is valid due to {\it special elastic properties of
%space-time.}
So, we can conclude, that within our approach, fundamental principles of
classical mechanics can be associated with the {\it special ($\sigma=1/2$)
elastic properties of space-time.}

Some other considerations  on physical nature of local hyperbolicity of
space-time and  role of boundary conditions in GR, relation between  $g$ and
$\Xi$ variational principles, some cosmological applications and 4D elastic
formulation of classical solids dynamics have been discussed in
\cite{kok2,kok3,kok4} as well . \vspace{0.5cm}

{\bf Acknowledgements.} I would like to express many thanks to prof. M.Pavsic
for useful discussions, E.P.Stern and (especially) K.Gudz for technical
assistance.

\end{document}